\documentstyle[aps,prl,preprint]{revtex}
\begin{document}
\draft

\title{Adsorption of colloidal particles in the presence of external
fields.}

\author{I. Pagonabarraga$^{(*)}$, J. Bafaluy$^{(**)}$ and J. M.
Rub\'{\i}$^{(*)}$\\ $^{(*)}$ Departament de F\'{\i}sica Fonamental,
Universitat de Barcelona\\ Av.  Diagonal 647, E-08028 Barcelona
(Spain).\\ $^{(**)}$ Departament de F\'{\i}sica, Universitat
Aut\`onoma de Barcelona\\ 08193 Bellaterra, Barcelona (Spain).}

\date{\today}

\maketitle

\begin{abstract}

We present a new class of sequential adsorption models in which the
adsorbing particles reach the surface following an inclined direction
({\it shadow models}).  Capillary electrophoresis, adsorption in the
presence of a shear or on an inclined substrate are physical
manifestations of these models.  Numerical simulations are carried out
to show how the new adsorption mechanisms are responsible for the
formation of more ordered adsorbed layers and have important
implications in the kinetics, in particular modifying the jamming
limit.

\end{abstract}

\setcounter{equation}{0}

\newpage

	In recent years much interest has been devoted to the study of
the adsorption of colloidal particles in solid surfaces\cite{1}
,\cite{2}.  In its complete formulation, the process becomes complex
because one should take into account the different forces which will
influence the rate of the arrival of particles to the surface, as well
as its relative position with respect to the preadsorbed
ones\cite{IPM},\cite{Baf}.  Recently, in order to study the adsorption
kinetics, attention has mainly been focused on the effect of surface
exclusion accounting for the fact that, once a particle is adsorbed,
it reduces the available surface for further adsorbing ones.  For
large particles like colloids, such a blocking effect is a non-linear
function of the surface coverage, since the excluded regions of
different particles may overlap each other.  Two major models have
been proposed to take these effects into account.  On one hand, in the
Random Sequential Adsorption (RSA) model\cite{Sch},\cite{Sch2}, if the
center of the incoming particle, whose position has been randomly
chosen on the surface, overlaps with a previously adsorbed one, it is
rejected and a new random position is selected; otherwise, the
particle is located irreversibly at that position.  Despite its
simplicity, experimental results on the adsorption of diffusing latex
spheres and polymers seem to agree with its
predictions\cite{Fed},\cite{Ram}.  On the other hand, in the Ballistic
Model (BM)\cite{Tal1},\cite{Tal2} if the center of the incoming
particle overlaps with a previously adsorbed one, it is allowed to
roll over it approaching the surface along the steepest descent path.
The particle is rejected only when it cannot reach the surface,
otherwise it is irreversibly located at that position as well.  This
model can describe the kinetics of the adsorption of large particles
when gravity is dominant\cite{Sch3}.

	We will introduce a new class of sequential adsorption models
in which particles adsorb according to RSA or BM rules, but reach the
surface along an inclined direction.  Our goal in this letter is to
show, through a numerical analysis, the new adsorption mechanisms with
respect to the ones proposed up to now, and their implications in the
different physical properties of the adsorbed layer.  In particular,
when considering BM rules, non-local effects will appear in the
dynamic of the incoming particles, which have important implications
in the values of physical quantities.  Moreover, this class of models
may cover different physical situations of interest.  It can either
correspond to the adsorption of particles in the presence of an
external field acting parallel to the surface or to the adsorption on
an inclined substrate.  Two specific examples of the first case are
capillary electrophoresis\cite{monning1994} and adsorption in the
presence of an imposed shear for sufficiently small diffusion
layers\cite{Ram}\cite{meinders1994}.  To be precise, we will study the
adsorption of disks of diameter one which arrive at the surface
forming an angle $\alpha$ with the normal to the line (see Fig.  1a).
As shown in the figure, an incident disk cannot land closer than a
distance of $1+\sigma=1/\cos \alpha$ to the right of a preadsorbed
particle, which can be thought of as though the adsorbed particle
projected a shadow on the line which cannot be occupied by an incident
particle, therefore, changing the overlapping mechanism with respect
to the standard models.  The different physical situations are
considered by studying the relationship between $\alpha$ (or
$\sigma$), with the corresponding physical parameters of interest.

	In order to find out more about the relevant features
introduced by these models, we have numerically analyzed the
adsorption kinetics of disks on a line of length 1000 diameters.  Two
major kinds of quantities are of interest in these models:  the
density of adsorbed particles as a function of time, and, in
particular, the maximum fraction of the surface covered by the
particles, referred to as the {\it jamming limit} $\rho_{\infty}$, and
the structure of the adsorbed phase, which follows from the pair
distribution function.  Those quantities will be studied in two models
keeping the rules of the standard RSA and BM.

	We will first consider the adsorption of hard disks following
RSA rules.  For this reason, disks are placed sequentially and
randomly on the line.  If the chosen position is at a distance smaller
than one to the left of a preadsorbed disk or at a distance smaller
than $1+\sigma$ to the right of a preadsorbed disk (see Fig.  1a), the
incoming particle is rejected and a new position is chosen; otherwise
it is irreversibly located at that position.  Therefore, the inclined
direction of arrival changes the overlapping mechanism with respect to
RSA.  It is worth noting that the relationship between the angle and
the excluded length makes it possible to consider the system as if we
had incoming particles of length one which, upon arrival at the
surface, deformed a length $1+\sigma$ towards the right.  This
asymmetry in the deformation makes these models different from the
restructuring-particle RSA proposed in the literature to study the
adsorption of certain proteins which deform at the surface\cite{TAR}.
In Fig.  2 we show the jamming limit as a function of $\sigma$
compared to its corresponding analytical expression

\begin{equation}
\rho(t)^{RSA}=\int_{0}^{t}
\exp\left[\int_0^\tau (-2+e^{-u}+e^{-(1+\sigma) u}) \frac{du}{u}
\right]  d\tau \label{eq1}
\end{equation}

\noindent by taking the limit $t\rightarrow \infty$, which has been
obtained from the corresponding integro-differential equations along
the same lines as in the standard RSA \cite{Sch2}.  An analysis of the
asymptotic behavior of eq.(\ref{eq1}) gives
$\rho_{\infty}^{RSA}-\rho(t)^{RSA}\sim 1/((1+\sigma)t)$ for long
times.  Details about the deduction of eq.(\ref{eq1}) will be
presented elsewhere.  As the incident angle, $\alpha$, increases, the
jamming decreases considerably since lengths between adsorbed disks
smaller than $1+\sigma$ cannot be occupied.  The structure of the
adsorbed disks is also modified.  As seen in Fig.  3a, the pair
distribution function at jamming exhibits two peaks located at $r=1$
and $r=1+\sigma$.  The former corresponds to the usual peak when
particles are in contact, whereas the latter results from the fact
that now $1+\sigma$ is a new minimum distance between particles.  A
careful theoretical analysis shows that both peaks are similar in
shape, therefore the second peak also diverges logarithmically at
jamming.

	In the second case, we will study the adsorption of disks
according to BM rules.  Again, positions are sequentially selected at
random.  If a chosen position overlaps disk 1 on the left hand side of
line $\Gamma$ (see Fig.  1b), the particle will roll to the left and
will end up on the line at contact with the preadsorbed disk.
However, if it is on the right hand side of that line at a distance
smaller than $1+\sigma$, it will end up at a separation $1+\sigma$ on
the surface.  Therefore, the probability that an incoming disk will
end at a length $1+\sigma$ to the right of a previously adsorbed one
is greater than the probability of landing at a distance 1 on its left
hand side.  This modified overlapping introduces significant changes
in the behavior of the incoming particle when there exists an
interaction with a second adsorbed disk.  In particular, new
restructuring effects appear with respect to standard BM.

	In fact, consider two particles at a distance
$\Delta<2+\sigma$, and an incoming disk overlapping the left one to
the right of line $\Gamma$ (see Fig.  1b).  When the disk comes into
contact with particle 1, it will roll over it trying to reach a
position at the surface at a distance $1+\sigma$ from it.  During its
motion, it can touch disk 2 before reaching the wall.  Then, if the
center of the incoming disk is to the left line $\Gamma '$, the
particle will try to reach the surface to the left of disk 2, if space
is available (Fig.  1b).  However, if its center is to the right of
that line, the incoming disk will try to reach the surface at a
distance $1+\sigma$ on the right hand side of disk 2 (Fig.  1c).  As a
result, the motion of the incoming particle will depend on a third
preadsorbed disk to the right of disk 2.  Therefore, the incoming disk
may jump over a number of adsorbed particles before reaching a place
at the surface.  This non-local adsorption mechanism cannot take place
in standard BM, since only when $\alpha \geq \pi /6$ may the center of
the incoming disk be to the right of $\Gamma'$ after rolling over
particle 1.  When $\sigma\geq 1$ ($\alpha\geq \pi /3$) the adsorption
process exhibits two new properties:  on one hand, there will be no
rejected particles before reaching the jamming because the incoming
disks will always jump over small gaps until they reach an available
space on the line.  On the other hand, when $\Delta = 1+\sigma$ , a
new singular contribution to the dynamics appears because all
particles rolling to the right over disk 1 will reach disk 2 at the
line $\Gamma'$.  Assuming that those disks then roll to its left, e.g.
due to gravity, they will end up at contact with disk 2 to the right
of disk 1.

	In Fig.  2 we have plotted the jamming density as a function
of $\alpha$.  At $\alpha=\pi /3$, a jump in the jamming density is
observed due to the singular contribution of gaps of length
$\Delta=1+\sigma$ in which now we may accommodate a new disk.  The
magnitude of the jump equals the fraction of disks which reach the
line following the new rolling mechanism explained before.  It is
possible to obtain an analytical expression for the deposition
according to BM rules, which takes the singular contribution into
account, but which neglects the non-linear adsorption mechanism
discussed previously.  For $\alpha< \pi/3$ it gives

\begin{equation}
\rho(t)^{BM}=\int_{0}^{t} d\tau \frac{d \rho^{RSA}(\tau)}{d \tau} (2
\tau+1) e^{-2 \tau} e^{2-e^{-\tau}-e^{-(1+\sigma)\tau}}
\label{bmrrt}
\end{equation}

 This analytic model is exact for $\alpha \leq \pi/6$, and gives an
accurate estimate of the jamming at larger angles, as shown in Fig.2.
This last feature explains the differences observed in the jamming
coverage near $\alpha=\pi/3$.  For $\alpha\geq\pi/3$, the coverage
increases linearly in time until jamming, since there are no rejected
particles, and when $\alpha\le \pi /3$, inspection of the analytical
result leads to the asymptotic time behavior
$\rho_{\infty}^{BM}-\rho(t)^{BM}\sim \exp(-t)/t$.

	In Fig.3b we show the pair distribution function at jamming.
The peaks observed correspond to clusters of particles, which
originate from the rolling mechanisms.  The separation between disks
in a given cluster is 1 or $1+\sigma$ (and also $\sigma-1$, if
$\sigma\geq 1$).  Therefore, the number of peaks increases at
distances in multiples of the diameter, since particles can be
distributed in two different sets of lengths which are incommensurable
when $\sigma$ is irrational.  For example, for $\sigma=.165$, we
observe two peaks near $r=1$, corresponding to separations $r=1$ and
$r=1+\sigma$, three peaks near $r=2$, corresponding to distances
$r=2$, $r=2+\sigma$ and $r=2 (1+\sigma)$, and so on.  The relative
areas under the peaks are equal to the relative probability for each
configuration of disks.  When $\sigma=1$, both kinds of distances are
commensurable, so when two particles are separated by $1+\sigma$,
there is exactly enough room for another particle, and therefore an
ordered sequence of peaks is obtained.  Moreover, for $\sigma \ge 1$
the number of peaks increases also due to the new singular
contribution.  It should be noted that the larger the value of
$\sigma$, the more ordered the structure of the adsorbed layer, the
reason being that when increasing $\sigma$, the probability of being
at a distance $1+\sigma$ increases relative to that at a distance 1
from a preadsorbed disk.

	In summary, we have introduced two new kinetic models ({\it
shadow models}) to study the adsorption of disks on a line taking into
account a driving in the direction parallel to the surface.  New
adsorption mechanisms which may take place have been shown, as well as
their effects both in the jamming coverage and in the pair
distribution function.  In particular, when considering BM rules, we
have shown that for angles larger than $\pi /3$, incident disks may
roll over a number of preadsorbed particles before reaching the line,
which means that the kinetics have become non-local.  Moreover, for
both BM and RSA rules, this driving originates more ordered substrates
on the lines, as shown through the pair correlation functions.

\acknowledgements

	We would like to acknowledge Profs.  P.  Schaaf and D.
Bedeaux for fruitful discussions.  This work has been supported by the
EEC under grant SCI$^{*}$-CT91-0696 and by DGICYT (Spain), grant
PB92-0895 (IP, JMR), and grant PB90-0676 (JB).

\begin{figure*}

\caption{a)Trajectory of an incoming particle of diameter unity
forming an angle $\alpha$ with the normal to the adsorbing line, in
the presence of two adsorbed disks,1 and 2, at a distance $\Delta$.
$1+\sigma$ represents the minimum distance between a preadsorbed disk
and the incoming particle on the line to its right.  b) and c) New
adsorption mechanisms in BM.  The incident disk rolls over disk 1 and
touches disk 2.  b) If its center is to the left of line $\Gamma'$ it
rolls to its left as in standard BM; c) if its center is to the right
of line $\Gamma'$ it will reach the line to the right of disk 2.  Its
kinetics depends now on a third sphere on the line.}

\caption{Jamming limit as a function of $2 \alpha/\pi$  for  RSA
(- - -)  and  BM (---) numerical . Analytically, RSA (ooo) and
BM ($^{...}$).}

\caption{Function $g(r)$ for a) RSA at $\sigma$= 0.04 ($^{...}$),
0.428(---) and 1.015 (- - -). b) BM at $\sigma$=0.165 (---) and 1.015
(- - -).}

\end{figure*}

\end{document}